\documentclass[10pt,conference]{IEEEtran}
\usepackage{multicol}

\usepackage{color}
\usepackage{soul}
\usepackage{amsmath}
\usepackage{graphics}
\usepackage{setspace}
\usepackage{cite}
\usepackage{latexsym}
\usepackage{float}
\usepackage{epsfig}
\usepackage{multirow}
\usepackage[table,xcdraw]{xcolor}
\usepackage{cite,cases,url}
\usepackage{amssymb}
\usepackage{graphicx}
\usepackage{epstopdf}
\usepackage{balance}
\usepackage{subcaption}
\usepackage{enumerate}
\usepackage{array}
\usepackage{algorithmic}
\usepackage{algorithm}
\usepackage{enumitem}
\usepackage[normalem]{ulem}
\usepackage{longtable}
\usepackage{ragged2e}
\usepackage{booktabs}
\usepackage{tabularx}
\usepackage{tikz}
\newcommand*\circled[1]{\tikz[baseline=(char.base)]{
            \node[shape=circle,draw,inner sep=0.5pt] (char) {#1};}}
            
\useunder{\uline}{\ul}{}

\newcolumntype{L}{>{\centering\arraybackslash}m{5cm}}
\newcolumntype{K}{>{\centering\arraybackslash}m{6cm}}
\newcolumntype{P}{>{\centering\arraybackslash}m{2.3cm}}
\newcolumntype{M}{>{\raggedright\arraybackslash}m{2cm}}
\newcolumntype{N}{>{\raggedright\arraybackslash}m{2.5cm}}

\begin{document}
\IEEEoverridecommandlockouts
\IEEEpubid{\makebox[\columnwidth]{979-8-3315-2037-3/25/\.00 \copyright 2025 IEEE \hfill} \hspace{\columnsep}\makebox[\columnwidth]{ }}
\title{%Next Generation Cellular Networks
The Role of Generative Artificial Intelligence in  Testing and Automation of Open RAN Systems:  GNG-OTA Framework }
%\hl{\hl{Myths and Facts of} }
\author{
\IEEEauthorblockN{ Aly Sabri Abdalla and Vuk Marojevic \\
} 
\normalsize\IEEEauthorblockA{Department of Electrical and Computer Engineering,  Mississippi State University, USA} 
\normalsize\IEEEauthorblockA{\{asa298,vm602\}@msstate.edu}
}
\maketitle
%TC:ignore 
\begin{abstract}
Open Radio Access Networks (O-RAN) empower operators with choice and customization through their disaggregated and interoperable architecture. However, ensuring interoperability in O-RAN presents challenges due to complex testing environments and evolving specifications. This paper addresses these challenges by proposing the Generative AI and Graph Neural Network-based Open RAN Testing and Automation (GNG-OTA) framework. GNG-OTA utilizes state-of-the-art AI techniques including generative AI, natural language processing, and graph neural networks to enable accurate, reliable, efficient, and automatic O-RAN interoperability and performance testing.
\end{abstract}
\vspace{-5 pt}
%TC:endignore
\IEEEpeerreviewmaketitle
\begin{IEEEkeywords}
O-RAN, testing, automation, AI, interoperability, performance, conformance, GAI, NLP, GNN.
\end{IEEEkeywords}

\section{Introduction}
\label{sec:intro}

Disaggregated and interoperable wireless network architectures offer significant advantages over traditional, vertically integrated approaches. This paradigm shift fosters a more competitive landscape, empowering network operators to select best-of-breed solutions from a diverse vendor ecosystem. Open interfaces and standardized protocols facilitate seamless integration of heterogeneous network components, enabling operators to tailor their infrastructure to specific deployment requirements. This disaggregation fosters technology agnosticism, allowing operators to advance beyond vendor lock-in and proprietary end-to-end solutions. The Open Radio Access Network (O-RAN) paradigm, an industry-driven initiative, provides a concrete framework for realizing these benefits through the strategic implementation of virtualized network elements (VNEs) and rigorously defined interfaces~\cite{9839628}, which is anticipated to drive down capital expenditures (CapEx) and operational expenditures (OpEx) for both operators and users.

To foster the widespread adoption of Open RAN solutions and ensure the success of open and interoperable wireless networks, a robust testing and evaluation ecosystem is crucial. This ecosystem should encompass two key pillars: 1) Development of standardized and industry-accepted test methodologies and 2) An open and collaborative environment for rigorous evaluation of Open RAN components and systems. This includes methodologies for ensuring seamless functionality between components from different vendors; evaluating key performance metrics like throughput, latency, and packet loss under various network conditions; assessing the vulnerability of Open RAN systems to cyberattacks; and ensuring compliance with security standards. For interoperability testing and evaluation in Open RAN architecture, several unique challenges stemming from the very nature of open and disaggregated networks must be addressed to unlock the full potential of Open RAN and foster a more competitive and innovative wireless ecosystem. These challenges include the increased complexity of the testing environment and tools due to the involvement of diverse hardware and software vendors to articulate a unifying open RAN architecture, unlike the traditional, vertically integrated networks from a single vendor. Additionally, the existence of a multitude of vendors and network configurations within open RAN networks results in a scenario where it would be a challenge to establish complete test visibility into all potential interoperability use cases. Moreover, the frequent ongoing efforts by the O-RAN Alliance and updates of the open RAN specifications to address emerging requirements would necessitate parallel adaptations to interoperability tests at the same speed to ensure they remain relevant and aligned with the latest standards~\cite{10757739}.

There have been some efforts recently through academia and O-RAN Alliance standard group to address these challenges including~\cite{hoschele20225g, yang2022interoperability, lin2022conformance}. In~\cite{hoschele20225g}, the authors discuss an experimental platform as a general O-RAN testbed to be used for benchmarking traditional and new vendors, facilitating examinations of O-RAN interoperability composed of OpenAirInterface (OAI) with the 5G-Core implemented over an industrial testbed. The work presented in~\cite{yang2022interoperability} presented an interoperability
test tool based on the Operations and Maintenance (OAM) interface specifications with the capability of achieving nearly 2.5 times less than manual testing. In~\cite{lin2022conformance}, authors propose using Multi-access Edge Computing (MEC) between the core network and RAN as a testing tool for validation of security and conformance performance of new devices attached to the network. There have been a few studies, such as the work presented in~\cite{ 10.1145/3629141} that discuss the use of Explainable AI for testing and validation of the deep reinforcement learning models deployed within the radio intelligent controllers (RICs) of the O-RAN.
However, none of these work items have discussed the possibility of utilizing advanced artificial intelligence (AI) techniques for the testing and automation of O-RAN interoperability and performance.  Therefore, in this paper, we survey the possible state-of-the-art AI techniques and propose \textbf{G}enerative \textbf{N}atural language \textbf{G}raph-based \textbf{O}pen RAN \textbf{T}esting and \textbf{A}utomation (GNG-OTA) framework. The GNG-OTA is an integrated state-of-the-art AI tester that can be deployed to enable accurate, reliable, efficient, and automatic O-RAN interoperability and performance testing ecosystem. 

%and  proposes [briefly introduce your solution/approach] to overcome this challenge and contribute to a more robust Open RAN interoperability testing ecosystem.}

Section II introduces the requirements and different test designs for O-RAN. Section III explores the potential of different state-of-the-art AI techniques for O-RAN testing. Section IV provides a detailed description of the proposed GNG-OTA framework based on the techniques explored in Section III for O-RAN testing. Section V concludes the paper.

\section{%Research
Requirements and Design of O-RAN Testing} %and Automation}
\label{sec:architecture}
\subsection{Near-RT and Non-RT RIC Testing}
% \textbf{Interoperability Testing:}
 Interoperability testing for both Near-RT and Non-RT RIC is crucial, as it ensures seamless integration of multi-vendor components within the O-RAN framework. This testing verifies that different RIC implementations can communicate and function %together 
 without issues %, thus fostering a cohesive and 
 for an efficient RAN ecosystem.

Interoperability testing for the A1-P and A1-EI interfaces is essential to ensure seamless integration and functionality of O-RAN components from different vendors. For A1-P, test scenarios include querying policy types, creating and deleting policy types, retrieving and updating policies, and notifying policy status changes~\cite{polese2023understanding}. These tests validate that policy information is accurately exchanged and interpreted across different RIC implementations. For A1-EI, the focus is on querying EI types, creating and managing EI jobs, and delivering EI job results. Test scenarios encompass querying EI job details, updating and deleting EI jobs, and ensuring accurate reporting of EI job statuses\cite{polese2023understanding}. These tests ensure that EI job information is correctly processed and communicated between diverse systems. Collectively, these interoperability tests are critical for maintaining a robust and cohesive O-RAN ecosystem, facilitating efficient and reliable network operations across multi-vendor environments.

The Near-RT and Non-RT RIC conformance testing is essential to ensure that O-RAN components adhere to specified standards and perform reliably under varied conditions. For Near-RT RIC, conformance testing focuses on low-latency operations and real-time control, validating functionalities such as querying policy types, creating and deleting policies, and receiving policy status notifications~\cite{huang2023validation}. This involves rigorous checks on message contents, structure, and response accuracy. Non-RT RIC testing %, on the other hand, 
emphasizes network optimization and orchestration using AI and ML, with test cases covering policy management, EI job handling, and notification delivery.

\subsection{O-DU and O-RU Testing}

Interoperability testing for O-DU and O-RU from different vendors is essential to ensure seamless operation within the O-RAN architecture. These tests focus on the O-RAN fronthaul interface, encompassing the M-Plane, S-Plane, C-Plane, and U-Plane. Each plane can be tested independently; however, comprehensive tests often require simultaneous activity across multiple planes, such as when devices are brought into service, and calls are established. For instance, M-Plane tests validate the startup sequence, ensuring the O-RU correctly interfaces with DHCP servers and NETCONF clients~\cite{lee2020design}. S-Plane tests verify synchronization, ensuring that the O-RU and O-DU maintain accurate frequency and time synchronization, even under varying temperature and traffic load conditions. The goal is to prove that the end-to-end functionality between the O-DU and O-RU meets the O-RAN fronthaul %CUS-Plane and M-Plane 
specifications.

C-Plane and U-Plane interoperability tests are crucial for validating radio system functionalities and multi-vendor interoperability. These tests ensure that User Equipment (UE) can establish C-Plane connections and perform initial U-Plane data transfers through the network, including O-DU and O-RU units from different vendors. Tests measure key metrics such as availability, accessibility, retainability, mobility, and data integrity. Customized interoperability testing configurations may be necessary when standard test profiles cannot be applied due to equipment limitations or licensing constraints, ensuring flexibility in testing scenarios. Overall, these tests ensure the correct timing and integrity of control and user data packets, verifying that the O-RU processing time and fronthaul latency meet the minimum feasible requirements.

Conformance testing for O-DU and O-RU units ensures that these components meet the necessary standards and specifications for seamless integration and operation within the O-RAN architecture. This testing encompasses various critical aspects, including session establishment with VLANs and DHCPv4, handling server subscriptions to notifications, managing connection supervision processes, and executing general commands and alarm notifications accurately~\cite{lake2021softwarization}. Additionally, user management and access control mechanisms are validated to ensure proper security implementations, while carrier deactivation and energy-saving configurations are tested for energy efficiency. Energy, Power, and Environmental (EPE) measurements are activated and deactivated to monitor the network's environmental impact. Uplink operations are verified against predefined 3GPP Reference Sensitivity levels, ensuring reliable communication links. Synchronization of the O-DU to incoming S-Plane timing information and correct. These tests confirm that the O-DU and O-RU units can operate reliably and efficiently, maintaining high performance and interoperability within a multi-vendor O-RAN environment.

\subsection{Other Performance Testing}

Other performance test cases were designed to validate various operational aspects of O-RAN units and interfaces. These tests include verifying the cell setup when the O-DU and O-RU are started separately. The tests ensure the proper broadcast of the Master Information Block (MIB) and System Information Blocks (SIB1 and SIB2) by the O-DU. Additionally, the tests verify the UE attachment procedure using both long and short preambles, UE registration and de-registration procedures in an end-to-end environment, and the UE PDU session establishment. The tests also include the verification of the successful establishment of a link between the O-DU and the SMO using static configuration, and the yang capability model exchange via the O1 interface. Further, they ensure successful cell activation based on configurations received from the SMO for co-located O-RU and O-DU, handling of subscribed measurement counters, and the establishment of the E2 interface with near-RT RIC. Periodic reporting by the O-DU to the near-RT RIC and successful UE attachment with given SSB periodicity and sub-carrier spacing are also tested. %, ensuring comprehensive validation of the O-RAN components' performance and interoperability.

\section{%What 
\textcolor{black}{Role of Intelligence in O-RAN Testing 
}}
\label{sec:testing}
Leveraging advanced AI techniques for testing and automation within the O-RAN ecosystem offers numerous advantages, enhancing efficiency, accuracy, and adaptability. This section delves into the applicability and benefits of integrating Generative AI (GAI), Natural Language Processing (NLP), and Graph Neural Networks (GNNs)  in O-RAN testing and automation.
\vspace{-3 pt}
\subsection{Generative AI (GAI)}
GAI, particularly models such as Generative Adversarial Networks (GANs) and Variational Autoencoders (VAEs), can significantly enhance O-RAN testing by generating synthetic data that closely mirrors real-world networks. %conditions. 
This capability allows for comprehensive testing scenarios, including rare and extreme conditions that might be difficult and time-consuming to replicate in the real world. %scenario. 
Generative AI can create diverse traffic patterns, user behaviors, and interference scenarios with an option to scale to test large and complex networks, enabling %more 
robust and exhaustive O-RAN testing. %of O-RAN components. 

GANs and VAEs are particularly suited for these tasks due to their technical architectures and inherent characteristics. The iterative process between the generator, which creates synthetic data samples, and the discriminator, which evaluates the authenticity of these samples, neural networks of the GAN enable produce highly realistic and diverse data sets~\cite{karapantelakis2024generative}. VAEs are designed to encode input data into a latent space and then decode it back to the original space, effectively capturing the underlying distribution of the data. This capability allows VAEs to generate new, plausible variations of network conditions and user behaviors. By manipulating the latent variables, VAEs can produce a wide range of scenarios, from typical to extreme %, facilitating comprehensive testing of 
for O-RAN testing~\cite{karapantelakis2024generative}.

GANs and VAEs %models 
facilitate the emulation of high policy query loads and dynamic policy updates, crucial for testing the interoperability of the A1-P and A1-EI interfaces. This allows for comprehensive validation that policy information is accurately exchanged and interpreted across Near-RT and Non-RT RICs, thereby ensuring a cohesive and efficient O-RAN ecosystem. In the context of O-DU and O-RU interoperability testing, GANs and VAEs can generate synthetic data to simulate various startup sequences, thereby validating the O-RU M-Plane's interface with DHCP servers and NETCONF clients. These models can also emulate diverse user behaviors and traffic patterns to validate the UE's ability to establish C-Plane connections and perform U-Plane data transfers through the network, ensuring that the end-to-end %functionality between the 
O-DU and O-RU functionality meets the O-RAN fronthaul specifications. Moreover, GAI techniques can establish scenarios to test synchronization accuracy under varying conditions, ensuring proper frequency and time synchronization between the O-RU and O-DU. These AI models help validate that the O-RU and O-DU maintain accurate synchronization, even under fluctuating environmental conditions and traffic loads, by simulating realistic interference patterns and synchronization challenges.

\vspace{-5 pt}
\subsection{Natural Language Processing (NLP)}
NLP leverages its ability to interpret and process textual data such as configuration files, log messages, and test case descriptions, automating the parsing and analysis of extensive log files, and quickly identifying anomalies and performance issues. By utilizing NLP techniques, O-RAN testing can achieve higher levels of automation, accuracy, and efficiency, addressing some of the most complex challenges in the telecom industry. This has become possible via the unique characteristics of the NLP techniques, including text parsing and understanding, information extraction, pattern recognition and anomaly detection, and semantic analysis.

NLP can be employed to automatically generate and manage test cases based on textual descriptions of testing requirements and specifications. By parsing natural language documents, such as O-RAN standards and vendor documentation, NLP algorithms can extract relevant test scenarios, procedures, and parameters~\cite{zhou2024large}. This automation reduces the time and effort required for test case creation, ensuring comprehensive coverage of all specified requirements. During O-RAN testing, large volumes of logs are generated, capturing detailed information about the system's behavior and performance. NLP can be utilized to analyze these logs, identifying patterns and anomalies that might indicate issues such as performance bottlenecks, configuration errors, or security breaches. Techniques like text mining and sentiment analysis can uncover hidden insights, enabling faster identification and resolution of problems. NLP can assist in maintaining traceability between test cases and their corresponding requirements. By analyzing the textual content of requirements documents and mapping them to specific test cases, NLP ensures that all requirements are adequately tested. Additionally, NLP can verify compliance with regulatory standards and guidelines by comparing the textual content of test results with the specified criteria~\cite{moore2025integratedllmbasedintrusiondetection}. NLP algorithms that can be used for such use cases include Bidirectional Encoder Representations from Transformers (BERT)-based model can be used to parse and understand complex technical documents. Also, the Pre-trained Transformer (GPT) models can be used to generate detailed test reports, summarizing the test results~\cite{zhou2024large}. 
\vspace{-3 pt}
\subsection{Graph Neural Networks (GNNs)}

GNNs can model the relationships and interactions between network components, providing deeper insights into network behavior and performance. GNNs can model the complex interactions between different vendor O-RAN component implementations. GNNs can learn to predict how well these components will work together by graphs representing each unit as a node and the interactions are edges. After that, the GNN model could be trained over historical test outputs and interactions to learn the patterns of successful test procedures and results by extracting features such as latency, error rates, and configuration parameters~\cite{wang2024graph}. This involves using techniques such as Graph Convolutional Networks (GCNs) or Graph Attention Networks (GATs). Once trained, the GNN can be used to predict the behavior of the network under various configurations and test scenarios. It can verify the success of new tests by comparing the observed interactions with the learned patterns. Deviations from these patterns can be flagged for further investigation, indicating potential issues~\cite{wang2024graph}.

GNNs can generalize and predict potential issues for various test use cases, for instance, the O-DU and SMO, and the O-DU and the near-RT connection establishment. The GNNs model can validate that policy updates and notifications are correctly propagated and adhered to in real-time between Near-RT RIC and Non-RT RIC via the A1 interface. The UE attachment procedure within the O-RAN architecture can also, be represented by the GNNs with its interactions to ensure that the attachment procedure follows the expected patterns and identifies anomalies that might indicate issues such as registration failures or authentication problems.

\section{{\textcolor{black}{AI-Assisted O-RAN Testing Framwork}} %What O-RAN %Can and 
%Cannot Do%\hl{---Limited Capabilities/Features}
}
\label{sec:implementation}
\begin{figure*}[h]
  \centering
  \includegraphics[width=1.8\columnwidth]{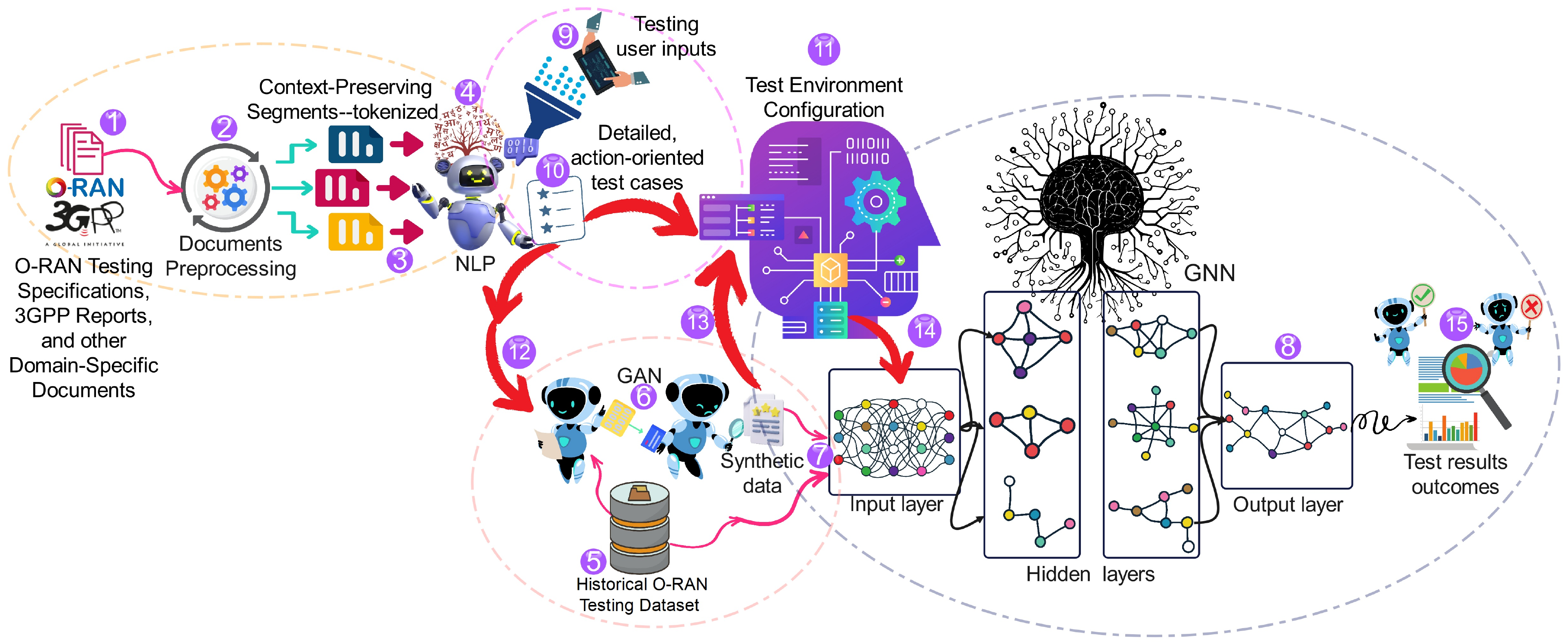}
\vspace{-6 pt}
  \centering \caption{GNG-OTA framework for O-RAN interpretability and performance testing and automation.} %State-of-the-art NLP, GAN, and GNN models are trained prior to performing O-RAN testing and automation, and later copies of these trained models are used for evaluating and assessing real-time test cases for different O-RAN components, interfaces, and functions. }
%   A robust teacher model is trained in the digital domain using DT, and a copy of the teacher is used for distillation for each student model. The location of the distillation training depends on the availability of system resources.
  \label{fig_framework}
%\textbf{ GNG-OTA Framework for O-RAN Testing and Automation.} State-of-the-art NLP, GAN, and GNN models are trained prior to performing O-RAN testing and automation, and later copies of these trained models are used for distillation for each student model. The training phase starts with NLP (steps \circled{1}-\circled{4}) after that the GAN is trained (steps \circled{5} and \circled{6}), and finally, the GNN is trained using historical and synthetic data (step \circled{7}). During the inference phase, the trained NLP provides detailed test case descriptions (step \circled{10}), the test environment is configured (step \circled{11}), trained GAN generates synthetic traffic and data (step \circled{13}), and trained GNN model the test observation logs and predict its behavior corresponding to previous pass tests (steps \circled{14} and \circled{15}).
\end{figure*}

In this section, we are proposing a detailed framework for testing O-RAN networks by using state-of-the-art intelligence techniques, particularly, GAN, NLP, and GNNs known as \textbf{G}enerative \textbf{N}atural language \textbf{G}raph-based \textbf{O}pen RAN \textbf{T}esting and \textbf{A}utomation (GNG-OTA). The framework is tailored to build a robust and adaptive environment for the system to perform tests with unique challenges in O-RAN networks' interoperability and performance %. %The framework envisioned for this O-RAN testing and automation %through advanced AI techniques 
%is 
as illustrated in Fig.~\ref{fig_framework}. 
% State-of-the-art NLP, GAN, and GNN models are trained prior to performing O-RAN testing and automation, and later copies of these trained models are used for evaluating and assessing real-time test cases for different O-RAN components, interfaces, and functions. The training phase starts with NLP (steps \circled{1}-\circled{4}) after that the GAN is trained (steps \circled{5} and \circled{6}), and finally, the GNN is trained using historical and synthetic data (step \circled{7}). During the inference phase, the trained NLP provides detailed test case descriptions (step \circled{10}), the test environment is configured (step \circled{11}), trained GAN generates synthetic traffic and data (step \circled{13}), and trained GNN model the test observation logs and predict its behavior corresponding to previous pass tests (steps \circled{14} and \circled{15})
\vspace{-3 pt}
\subsection{GNG-OTA Training Phase}
The first phase in the GNG-OTA framework consists of the NLP preprocessing of all the relevant O-RAN testing specifications and 3GPP reports, both standard and domain-specific, to generate a massive corpus (step \circled{1} in Fig.~\ref{fig_framework}) to be used as an input for the following preprocessing and refinement process of the framework. One level of preprocessing further refines this corpus by breaking documents up into context-preserving text segments (step \circled{2}). This ensures that each segment is self-contained and does not split crucial events across segments. After this pre-processing step, the preprocessed corpus and context-preserving segments are tokenized in the form of words or phrases that retain the context for the NLP training (step \circled{3}). A pre-trained NLP model can be utilized at this point, such as BERT or GPT to be fine-tuned via supervised learning with the annotated domain-specific terminology and context-preserving data (step \circled{4}). The NLP model is trained to recognize specific keywords and phrases related to test requirements, such as interoperability, latency, throughput, synchronization, and different O-RAN architecture units. After steps \circled{1} to \circled{4} shown in Fig.\ref{fig_framework} take place, the outcome would be a trained NLP model over various O-RAN and 3GPP testing procedures and reports that ensure that the NLP is capable of extracting key information from the high-level descriptions input data and converting it to detailed test requirements.

The second training phase entails training the GAN within the framework, historical O-RAN interactions between components during tests, O-RAN open interfaces traffic patterns, test events sequences, exchanged packets and messages, timings, and test outcomes results need to be collected as a training set for the GAN and GNN models (step \circled{5}). The generator and discriminator components within the GAN algorithm could be architected as Deep Convolutional GAN (DCGAN) or Wasserstein GAN (WGAN) according to the complexities and nature of the collected historical dataset. Over the iterative training process between the generator and discriminator of the GAN model (step \circled{6}) the generator will be able to learn how to artificially output highly realistic data patterns that closely mimic the real samples of real-world O-RAN tests.

The third phase is the training of the GNN to model the O-RAN components under test as a graph, capturing their interactions and dependencies during testing. The generated synthetic data and previously collected real samples are then fed to the GNN model in order to extract the relevant features corresponding to different test use cases (step \circled{7}). The historical real data collected and the GAN synthetic data including configurations, performance metrics, and outcomes, are used to construct this graph. The GNN will then model the O-RAN components under testing as a graph where the testing unit and its neighboring units are represented as nodes and its interactions with other units are illustrated as edges. Within the GNN, each component of the graph (nodes, edges, and the global context) is processed using separate multilayer perceptrons (MLPs), collectively referred to as a GNN layer. The MLP processes the node vector and outputs a learned node vector, which encapsulates the features and characteristics of that node after considering the interactions with its neighboring nodes. Similarly, for each edge in the graph, an MLP is applied to learn a per-edge embedding that represents the interaction between the connected nodes, capturing information such as latency, error rates, achieved rate, and configuration parameters. The entire graph is also represented by a global-context vector processed by the MLP to learn a single embedding for the entire graph, summarizing the overall state and context of the O-RAN network being tested (step \circled{8}).  This supervised learning helps the GNN recognize patterns of successful tests and identify any deviations from expected test case patterns by observing graph nodes and edge interactions, exchanged packets, and learning from features.

% For example, a user can enter test keywords, such as O-DU, O-CU, and interoperability. Based on that, the NLP model should generate multiple test use case descriptions including...... 

\subsection{Inference Phase}% and Workflow Example}
The process begins with the test requirements at a high level that are captured from the GAG-OTA user in a natural language (step \circled{9} of Fig.~\ref{fig_framework}). These requirements are then passed to the trained NLP module that captures the exact intent of the user and converts it into action-oriented test cases. The NLP's generated action-oriented test cases would contain a description of what needs to be tested, step-by-step test procedures, and specific parameters needed to finalize the test and monitored during the test phase (step \circled{10}). These outputs are then structured into a machine-readable format, such as JSON or XML file to be accessed or processed by the other test units of the framework. These JSON or XML test scenarios and requirements are then structured to be fed as input configuration files for setting up the O-RAN test environment for the specific units, interfaces, or functions under test (step \circled{11}). Additionally, these NLP output configuration files are streamed to the GAN (step \circled{12}) to provide synthetic traffic patterns during the inference phase based on the specified test case to simulate new or unseen data for evaluating the performance under test (step \circled{13}). The test is then executed using custom scripts for testing automation within the configured O-RAN test environment and logs of tests are collected and fed to GNN for nodes and edges graph construction with embedded logs and features collected during the test for providing an in-depth graph characterization (step \circled{14}). The GNN's role is to predict potential interoperability issues and performance bottlenecks by comparing the current test graphs output with historical test outputs related to under-test components, functions, or interfaces (step  \circled{15}).

% The following provides an example of the GNG-OTA process and flow.

% A user inputs a requirement such as, "Test the synchronization accuracy between O-RU and O-DU under varying traffic loads." The NLP module generates a detailed test case specifying traffic conditions and expected outcomes. The GAN creates synthetic traffic data to simulate these conditions. The GNN analyzes synchronization accuracy using the synthetic data, flags any deviations and ensures that the O-RAN components meet performance standards.

% By leveraging advanced AI techniques, this framework enhances the efficiency, accuracy, and comprehensiveness of O-RAN testing, facilitating the development and deployment of reliable, high-performance O-RAN networks. The integration of NLP, GANs, and GNNs ensures that the framework can handle the complexity and variability of O-RAN networks, providing a robust and adaptive solution for interoperability and performance testing.

%\section{Testbed Resources}
%\label{sec:testbed}
%\input{include/testbeds}

% \section{O-RAN Evolution: To Enable the Requested Features} %Architecture}
% \label{sec:contribution}
% \input{include/contribution.tex}

% \section{Research Directions to O-RAN}
% \label{sec:case}
% \input{include/directions.tex}

\section{Conclusions}
\label{sec:conclusions}
This paper introduced the GNG-OTA framework that leverages state-of-the-art AI techniques like GAI, NLP, and GNNs for automating and testing O-RAN interoperability and performance, addressing the challenges of complex testing environments and the need to adapt to evolving specifications. The framework's ability to preprocess extensive protocol specifications, generate diverse test cases, predict potential issues through graph-based analysis, and provide detailed result reporting addresses the challenges inherent in complex testing environments and the continual evolution of O-RAN specifications by significantly improving testing efficiency, reliability, and accuracy. By integrating these cutting-edge AI techniques, GNG-OTA not only streamlines the testing process but also ensures comprehensive coverage and deeper insights into network behavior.
%By significantly improving efficiency, reliability, and accuracy, GNG-OTA has the potential to pave the way for a more robust and thriving O-RAN ecosystem. 

\section*{Acknowledgement}
This work was supported by the National Telecommunications and Information Administration (NTIA) under Award No.~28‑60‑IF012, as well as the Office of Naval Research under Award No. N00014-23-1-2808.

\balance

\bibliographystyle{IEEEtran}
\bibliography{aly}

% \section*{Biographies}
% \footnotesize
% % \vspace{0.2cm}
% % \noindent
% % \textbf{Joshua Moore} (jjm702@msstate.edu)
% % is a PhD student in the Department of Electrical and Computer Engineering at Mississippi State University, Starkville, MS, USA. His research interests include O-RAN, 5G/next-G communications, and RAN Management and Orchestration.

% \vspace{0.2cm}
% \noindent
% \textbf{Aly Sabri Abdalla} (asa298@msstate.edu)
% is an Assistant Research Professor in the Department of Electrical and Computer Engineering at Mississippi State University, Starkville, MS, USA. His research interests are on wireless communication and networking, software radio, spectrum sharing, wireless testbeds and testing, and wireless security with application to mission-critical communications, open radio access network (O-RAN), unmanned aerial vehicles (UAVs), and reconfigurable intelligent surfaces (RISs).

% % \vspace{0.2cm}
% % \noindent
% % \textbf{Charles Ueltschey}
% % (cmu32@msstate.edu) is an undergraduate student in computer science at Mississippi State University, Starkville MS, USA. His research interests are Wireless communications, O-RAN security, and Macine Learning.

% \vspace{0.2cm}
% \noindent
% \textbf{Vuk Marojevic} (vuk.marojevic@msstate.edu) is an associate professor in electrical and computer engineering at Mississippi State University, Starkville, MS, USA. His research interests include resource management, vehicle-to-everything communications and wireless security with application to cellular communications, mission-critical networks, and unmanned aircraft systems.

\end{document}